\documentclass[11pt]{article}
\usepackage{amssymb}
\usepackage{epsfig}

\begin{document}


\def\IR{{\mathbb R}}
\def\IN{{\mathbb N}}
\def\IC{{\mathbb C}}
\def\epsi{\varepsilon}
\def\1I{\mathchoice{ \hbox{${\rm 1}\hspace{-2.3pt}{\rm l}$} }
                   { \hbox{${\rm 1}\hspace{-2.3pt}{\rm l}$} }
                   { \hbox{$ \scriptstyle  {\rm I}\!{\rm N}$} }
                   { \hbox{$ \scriptscriptstyle  {\rm I}\!{\rm N}$}}}

\def\kasten#1{\mathop{\mkern0.5\thinmuskip
                      \vbox{\hrule
                            \hbox{\vrule
                                  \hskip#1
                                  \vrule height#1 width 0pt
                                  \vrule}%
                            \hrule}%
                      \mkern0.5\thinmuskip}}

\def\qed{\mathchoice{\kasten{8pt}}%
                       {\kasten{6pt}}%
                       {\kasten{4pt}}%
                       {\kasten{3pt}}}%

\renewcommand{\labelenumi}{{\rm(\roman{enumi})}}

\def\epsi{\varepsilon}

\newcounter{mathe}
\newenvironment{mathe}[1]
{ \par\bigskip\refstepcounter{mathe}\noindent\textbf{#1 \arabic{mathe}.
 }\mdseries\ }
{\par\bigskip\mdseries}


\title{Adiabatic Decoupling and Time-Dependent Born-Oppenheimer Theory}
\author{Herbert Spohn and Stefan Teufel\\ \\
\em Zentrum Mathematik and Physik Department,\\
\em Technische Universit\"at M\"unchen,\\
\em 80290 M\"unchen, Germany\\
email: spohn@ma.tum.de, teufel@ma.tum.de}
\date{July 9, 2001}
 \maketitle
\begin{abstract}
We reconsider the time-dependent Born-Oppenheimer theory with the goal to carefully separate 
between
the adiabatic decoupling of a given group of energy bands from their orthogonal subspace
and the semiclassics within the energy bands. Band crossings are allowed and our
results are local in the sense that they hold up to the first time when a band crossing is
encountered. The adiabatic decoupling leads to an effective Schr\"odinger equation
for the nuclei, including contributions from  the Berry connection.
\end{abstract}

\section{Introduction}

Molecules consist of light electrons, mass $m_{\rm e}$,  and heavy nuclei, mass $M$ which depends
on the type of nucleus.
Born and Oppenheimer \cite{BornOppenheimer}
wanted to explain some general features of molecular spectra and 
 realized that, since the ratio $m_{\rm e}/M$ is small, it could be used as an expansion parameter
for the energy levels of the molecular Hamiltonian. The time-{\em independent}
Born-Oppenheimer theory has been put on  firm mathematical grounds by Combes, Duclos, and Seiler
 \cite{CDS},  Hagedorn \cite{HagedornTI}, and more recently in \cite{KMSW}.

With the development of tailored state preparation and ultra precise time resolution there is a
growing interest in understanding and controlling the dynamics of molecules, which requires 
an analysis of the solutions to the time-{\em dependent} Schr\"odinger equation, again exploiting that
$m_{\rm e}/M$ is small. The molecular Hamiltonian is of the form
\begin{equation} \label{SBO1}
H = \frac{\hbar^2}{2m_{\rm e}}\Big(-i\nabla_x - A_{\rm ext}(x)\Big)^2  + \frac{\hbar^2}{2M} 
\Big(-i\nabla_X + A_{\rm ext}(X)\Big)^2
+ V_{\rm e}(x) + V_{\rm en}(X,x) + V_{\rm n}(X)\,.
\end{equation}
For notational simplicity we ignore spin degrees of freedom and assume 
that all nuclei have the same mass. We have $k$ electrons with positions
$\{x_1,\ldots,x_k\}=x$ and $l$ nuclei with positions $\{X_1,\ldots,X_l\}=X$.
The first and second term of $H$ are the kinetic energies 
of the electrons and of the nuclei, respectively. An external magnetic field
is  included through the vector potential $A_{\rm ext}$. 
Electrons and nuclei interact via the static Coulomb potential. Therefore $V_{\rm e}$ is
the electronic, $V_{\rm n}$ the nucleonic repulsion, and $V_{\rm en}$ the 
attraction between electrons and nuclei. $V_{\rm e}$ and $V_{\rm n}$ may also contain 
an external electrostatic potential.

In atomic units ($m_{\rm e}=\hbar=1$) the Hamiltonian (\ref{SBO1}) can be written more concisely as
\begin{equation} \label{SBO21}
H  =   \,\frac{m_{\rm e}}{M}\,\frac{1}{2}
\Big(-i\nabla_X + A_{\rm ext}(X)\Big)^2 + H_{\rm e}(X)\,,
\end{equation}
emphasizing that the nuclear kinetic energy will be treated as a ``small perturbation''.
$H_{\rm e}(X)$ is the electronic Hamiltonian for given position $X$ of the nuclei,
\begin{equation} \label{SBO3}
H_{\rm e}(X) =  \frac{1}{2} \Big(-i\nabla_x - A_{\rm ext}(x)\Big)^2 
 + V_{\rm e}(x) + V_{\rm en}(X,x) + V_{\rm n}(X)\,.
\end{equation} 
$H_{\rm e}(X)$ is a self-adjoint operator on the electronic Hilbert space $L^2(\IR^{3k})$
restricted to its antisymmetric subspace.
Later on we will need some smoothness
of $H_{\rm e}(X)$, which can be established easily if the electrons are
treated as point-like and the nuclei have an extended, rigid charge distribution.

Generically $H_{\rm e}(X)$ has, possibly degenerate, eigenvalues $E_1(X)<E_2(X)<\ldots$
which  terminate at the continuum edge $\Sigma(X)$. Thereby one obtains the band structure
as plotted schematically in Figure 1.
The discrete bands $E_j(X)$ may cross and possibly merge into the continuous spectrum as indicated
in Figure 2.
\begin{figure} \label{F1}\begin{center}
\epsfig{file=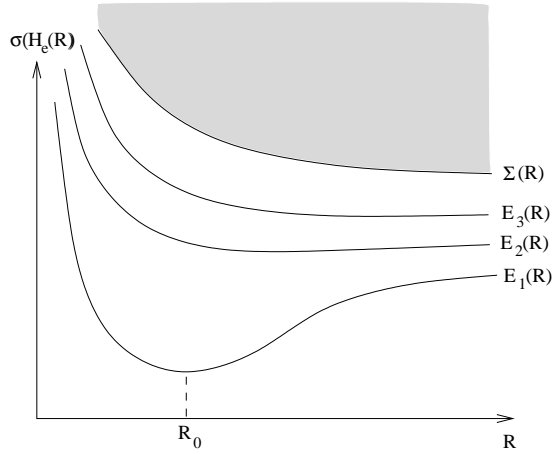, height=6cm, scale=1} \end{center}
\caption{The schematic spectrum of $H_{\rm e}(R)$ for a 
diatomic molecule as a function of the separation $R$ of the two nuclei.}
\end{figure}

Comparing kinetic energies, we find for the speeds $|v_{\rm n}|\approx(m_{\rm e}/M)^{1/2}|v_{\rm e}|$,
which means that on the atomic scale the nuclei move very slowly.
If  we regard $X(t)$ as a given nucleonic
trajectory, then $H_{\rm e}(X(t))$ is a Hamiltonian with slow time variation
and  the time-adiabatic theorem \cite{Kato,JoyePfister,AvronElgart} can be applied \cite{Bornemann}.
For us $X$ are quantum mechanical degrees of freedom. The Hamiltonian
$H$ of (\ref{SBO21}) is time-independent and we can only exploit that the nucleonic
Laplacian carries a small prefactor. To distinguish, we refer to our situation
as space-adiabatic. Since the nuclei move very slowly,  their dynamics must be followed
over sufficiently long times. From the speed ratio we conclude that 
these times are of order $(m_{\rm e}/M)^{1/2}$ in atomic units. To simplify notation
we define
\begin{equation}\label{epsdef}
\epsi = \sqrt{\frac{m_{\rm e}}{M}}
\end{equation}
as the small dimensionless parameter. Then
\begin{equation} \label{SBO2}
H^\epsi  =    \epsi^2\frac{1}{2}\Big(-i\nabla_X + A_{\rm ext}(X)\Big)^2  + H_{\rm e}(X)\,,
\end{equation}
and we want to study the solutions of the time-dependent Schr\"odinger equation
\begin{equation} \label{SE2}
i\epsi\frac{\partial\psi}{\partial t} = H^\epsi \psi
\end{equation}
in the limit of small $\epsi$.

The crude physical picture underlying the analysis of (\ref{SE2}) is that the nuclei
behave semiclassically because of their large mass and that the electrons rapidly  adjust to the 
slow nucleonic motion. Thus, in fact, the time-dependent Born-Oppenheimer approximation 
involves two limits. If the electrons are initially in the eigenstate 
$\chi_j(X_0)$ of the $j$-th band with energy $E_j(X_0)$, where $X_0$ is the approximate
initial configuration of the nuclei, then the $j$-th
band is adiabatically protected provided there is an energy gap separating it
from the rest of the spectrum. Thus at later times, up to small error, the electronic wave function
is still in the subspace corresponding to the $j$-th band. But this implies that the nuclei
are governed by the Born-Oppenheimer Hamiltonian
\begin{equation}\label{BOHi}
H^\epsi_{\rm BO} =   \epsi^2\frac{1}{2} \Big(-i\nabla_X + A_{\rm ext}(X)\Big)^2 + E_j (X)\,.
\end{equation} 
Since $\epsi\ll 1$, $H_{\rm BO}^\epsi$ can be analyzed through semiclassical methods
where to leading order the contributions come from the classical flow $\Phi^t$
corresponding to the classical Hamiltonian  
$H_{\rm BO}^{\rm cl}= \frac{1}{2}p^2+ E_j(q)$ on nucleonic phase space.

In general, $E_j(X)$ may touch another band as $X$ varies.
To allow for such  band crossings we introduce the region $\Lambda\subset\IR^n$, $n=3l$, in nucleonic 
configuration space, such that $E_j$ restricted to $\Lambda$ does not cross or touch any other
energy band. The classical flow $\Phi^t$ then has $\Lambda\times \IR^n$ as phase space and is
 defined only up to the time when it first hits
the boundary $\partial \Lambda\times\IR^n$. Up to that time (\ref{BOHi}) still correctly
describes the quantum evolution. To follow the tunneling through a band crossing other 
methods have to be used \cite{HagedornCross,Gerard}, in particular, the codimension of the crossing
is of relevance.

The mathematical investigation of the time-dependent Born-Oppenheimer theory
was initiated and carried out in great detail by Hagedorn. In his pioneering work 
\cite{Hagedorn80} he constructs approximate solutions to (\ref{SE2}) of the form
$\phi_{q(t),p(t)}\otimes\chi_j(q(t))$, where $\phi_{q(t),p(t)}$ is a coherent state
carried along the classical flow, $(q(t),p(t)) = \Phi^t(q_0,p_0)$. The difference to the
true solution with the same initial condition is of order $\sqrt{\epsi}$ in the
$L^2$-norm over times of order $\epsi^{-1}$ in atomic units and the approximation holds
until the first hitting time of $\partial \Lambda\times\IR^n$. In a recent work Hagedorn and Joye
\cite{Hagedorn00} construct solutions to (\ref{SE2}) satisfying exponentially small error
estimates. In Hagedorn's approach the ``adiabatic and
semiclassical limits are being taken simultaneously, and they are coupled 
\cite{Hagedorn00}''.

In our paper we carefully separate the space-adiabatic and the semiclassical limit. One 
immediate benefit is the generalization of the first order analysis of Hagedorn from coherent
states to arbitrary wave functions.

Let us explain our result for the space-adiabatic part in more detail. We assume that there is
some region $\Lambda\subset\IR^n$ in the nucleonic configuration space, such that some
subset $\sigma_*(X)$ of $\sigma(H_{\rm e}(X))$ is separated from the 
remainder of the spectrum by a gap for all $X\in\Lambda$, i.e.\
\[
{\rm dist}\big(\sigma_*(X), \sigma(H_{\rm e}(X))\setminus\sigma_*(X)\big)\geq d>0\quad 
{\rm for\,\,all\,\,}X\in\Lambda\,.
\]
$\Lambda$ could be punctured by small balls (for $n=2$) because of band crossings. 
$\Lambda$ could also terminate because the point spectrum merges
in the continuum, which physically means that the molecule loses an electron through
ionization. 
Let $P_*(X)$ be the spectral projection  of $H_{\rm e}(X)$ associated with $\sigma_*(X)$
and $P_* = \int^\oplus_\Lambda\,dX\,P_*(X)$. We will establish that the unitary time evolution 
$e^{-iH^\epsi t/\epsi}$ agrees on Ran$P_*$ with the diagonal 
evolution $e^{-iH_{\rm diag}^\epsi t/\epsi}$ generated by 
$H_{\rm diag}^\epsi:= P_*H^\epsi P_*$
up to errors of order $\epsi$ as long as the leaking 
through the boundary of $\Lambda$ is sufficiently small. 

To complete the analysis one has to control the flow of the wave function
through $\partial\Lambda$. One possibility is to simply avoid the problem by  assuming that 
$\Lambda = \IR^n$, hence $\partial \Lambda =\emptyset$. We will refer to this case as
 a {\em globally isolated} band.
Of course, the set $\{(X,y)\in \IR^n\times\IR: y\in \sigma_*(X)\}$ may
 contain arbitrary band crossings. As one of our main results,
 we prove that the subspace
Ran$P_*$ is adiabatically protected. In particular for the purpose of 
studying band crossings the full molecular Hamiltonian may be replaced by a simplified
model with two bands only. 

In general one has $\partial \Lambda\not=\emptyset$, to which we refer as 
a {\em locally isolated} band. 
To estimate the flow out of $\Lambda$ the only technique available seems to be
semiclassical analysis. But this requires a control over the semiclassical
evolution, for which one needs, at present, that
$\{(X,y)\in \Lambda\times\IR: y\in \sigma_*(X)\}$ contains no band crossings.
Then $\{(X,y)\in \Lambda\times\IR: y\in \sigma_*(X)\} = 
\cup_j\,\{(X,y)\in\Lambda\times\IR: y =  E_j (X)\}$ is the disjoint union of possibly
degenerate  energy bands $E_j(X)$. We will prove that 
 each band separately is adiabatically protected.
\begin{figure} \label{F2}\begin{center}
\epsfig{file=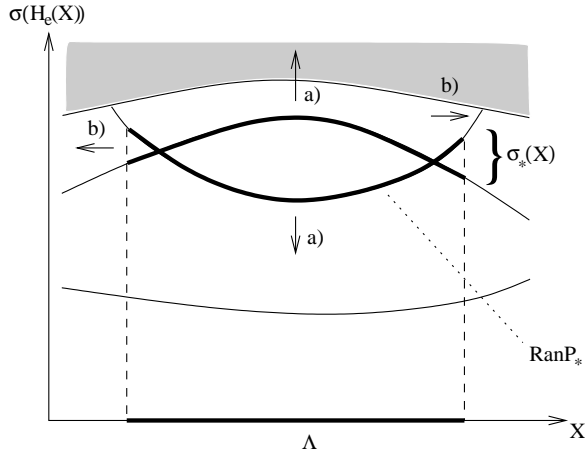, height=6cm, scale=1} \end{center}\caption{The wave function can leave
Ran$P_*$ in two different ways. Either by transitions to other bands (a) or through the
boundary of $\Lambda$ (b).}
\end{figure}

In the special case where $\sigma_*(X)=E_j(X)$ is a nondegenerate eigenvalue for $X\in\Lambda$,
$e^{-iH_{\rm diag}^\epsi t/\epsi}$ is well approximated through
$e^{-i H_{\rm BO}^\epsi t/\epsi}$  on $L^2(\IR^{n})$. Since $H_{\rm BO}^\epsi$
is a standard semiclassical operator, one can easily control the
$X$-support of the wave function and therefore prove a result for rather general 
$\Lambda\subset\IR^{n}$, for the details see Theorem \ref{MT}.
Roughly speaking, it says that
if $\phi_{t}$ is a solution of the effective
 Schr\"odinger equation for the nuclei
\begin{equation} \label{BOSE}
i\epsi\frac{\partial\phi_{t}}{\partial t} = 
H^\epsi_{\rm BO} \phi_{t}\,,
\end{equation}
with supp$\phi_0\subset\Lambda$, 
then, modulo an error of order $\epsi$, 
\[
\psi_{t} :=\phi_{t}(X) \chi_j(X,x)
\]
is a solution of the full Schr\"odinger equation (\ref{SE2})
with initial condition $\psi_0(X,x) = \phi_0(X)\chi_j(X,x)$ 
as long as $\phi_t$ is supported in $\Lambda$ up to $L^2$-mass of order $\epsi$.
This maximal time span can be
computed using the classical flow $\Phi^t$.

As first observed by Mead and Truhlar \cite{Mead}, in general 
$H_{\rm BO}^\epsi$  acquires as a first order correction an additional vector potential 
$A_{\rm geo}(X) =-i\langle \chi_j(X),\nabla_X \chi_j(X)\rangle$
and (\ref{BOHi}) has to be replaced by
\begin{equation}\label{BOHii}
H^\epsi_{\rm BO} =   \epsi^2\frac{1}{2} \Big(-i\nabla_X + A_{\rm ext}(X)+A_{\rm geo}(X)
\Big)^2 + E_j (X)\,.
\end{equation} 
Multiplying  $\chi_j(X)$ with a smooth $X$-dependent phase factor
induces a gauge transformation for $A_{\rm geo}$, which implies that   
the physical predictions based on (\ref{BOHii}) do not change, as it should be. 
As noticed in \cite{Mead}, if $\Lambda$ is not contractible, then $A_{\rm geo}$ cannot
be removed through a gauge transformation and (\ref{BOHii}) and (\ref{BOHi})
 describe different physics.
Berry  realized
that  geometric phases appear whenever the Hamiltonian has slowly changing parameters.
Therefore $A_{\rm geo}(X)$  is referred to as Berry connection, cf.\ \cite{Berryphase}
for an instructive collection of reprints. 
In fact, the motion of nuclei as governed by the Born-Oppenheimer Hamiltonian (\ref{BOHii})
is one of the paradigmatic examples for geometric phases.

If  $\sigma_*(X) = E(X)$ is  $k$-fold degenerate, not much of the above analysis changes.
$H_{\rm BO}^\epsi$ becomes  matrix-valued and acts on $L^2(\IR^n)^{\oplus k}$, i.e.\
\[
H_{\rm BO}^\epsi = \left(  \frac{\epsi^2}{2}  \Big(-i\nabla_X + A_{\rm ext}(X)
\Big)^2  + E_j (X)\right)\,{\bf 1}_{k\times k}
+ \frac{\epsi}{2}\Big( (-i\epsi\nabla_X) \cdot A_{\rm geo}(X) + A_{\rm geo}(X)\cdot 
(-i \epsi\nabla_X)\Big)\,.
\]
The  connection $A_{\rm geo}(X)$ contains in general also off-diagonal terms and 
matrix-valued semiclassics must be applied. 
However, since the only nondiagonal term is in the subprincipal symbol, the leading order 
semiclassical analysis reduces to the scalar case and, in particular, agrees with the 
nondegenerate band case. We do not carry out the straightforward  extension of
 Theorem \ref{MT} below to the degenerate band case,
because the technicalities of matrix-valued semiclassics would obscure 
the simple ideas behind our analysis.

In their recent work \cite{MS} Martinez and Sordoni independently study 
the time-dependent Born-Oppenheimer approximation as based on 
 techniques developed by Nenciu and Sordoni \cite{NS}. 
  They consider the case of a globally isolated band for a Hamiltonian of the form (\ref{SBO1}) 
 with smooth $V$ and $A_{\rm ext}=0$.
They succeed in proving the
adiabatic decoupling to any order in $\epsi$ for  subspaces
$P_*^\epsi$ which are $\epsi$-close to the unperturbed subspaces $P_*$  considered by us.
  With this result, in principle, 
 higher order corrections to the effective 
Hamiltonian (\ref{BOHi}) could be computed. 

The paper is organized as follows. Section 2 contains the precise formulation 
of the results.
Section 3 gives  a short discussion  
of the semiclassical limit of $H_{\rm BO}^\epsi$ and on how such results extend 
 to the full molecular system. 
Proofs are provided in Section 4. In spirit they rely on techniques developed
in \cite{SpohnT} in the context of the semiclassical limit for dressed electron states.
In practice the Born-Oppenheimer approximation requires several novel constructions,
since the ``perturbation'' $-\frac{\epsi^2}{2}\Delta$ increases quadratically.

Our results can be formulated and proved in a more general framework 
dealing with, possibly time-dependent, perturbations of fibered operators. 
Also the gap condition can be removed by using arguments similar to those 
developed by Avron and Elgart in \cite{AvronElgart}. The general operator theoretical
 results will appear elsewhere \cite{general}.

\section{Main results}

The specific form (\ref{SBO3}) of the electronic part of the  Hamiltonian 
will be of no importance in the following.
Thus we only assume that 
\[
H_{\rm e} = \int^\oplus_{\IR^n}\,dX\, H_{\rm e}(X)\,,\qquad H_{\rm e}(X) = H_{\rm e0} + 
H_{\rm e1}(X)\,,
\]
where $H_{\rm e0}$ is self-adjoint on some dense domain $\mathcal D\subset \mathcal{H}_{\rm e}$ 
and bounded from below and $H_{\rm e1}(X)\in \mathcal{L}(\mathcal{H}_{\rm e})$
is a continuous family of self-adjoint operators, bounded uniformly for $X\in \IR^n$. Thus
$H_{\rm e}$ 
is self-adjoint on $D(H_{\rm e}) = L^2(\IR^n)\otimes \mathcal{D}\subset\mathcal{H}:=
L^2(\IR^n)\otimes \mathcal{H}_{\rm e} $ and bounded from below.
For the definition of $L^2(\IR^n)\otimes \mathcal{D}$ we equip $\mathcal D$ 
with  the graph-norm $\|\cdot\|_{H_{\rm e0}}$, i.e., for $\psi\in\mathcal{D}$,
$\|\psi\|_{H_{\rm e0}} = \|H_{\rm e0} \psi\|+\|\psi\|$.

Let  $A_{\rm ext}\in C^1_{\rm b}(\IR^n,\IR^n)$, where
for any open set $\Omega\subset\IR^m$, $m\in\IN$, 
 $C^k_{\rm b}(\Omega)$ denotes the set of functions $f\in C^k(\Omega)$ such that 
for each multi-index $\alpha$ with $|\alpha|\leq k$ there exists a $C_{\alpha}<\infty$ with
\[
\sup_{x\in\Omega} |\partial^\alpha f(x)|\leq C_{\alpha}\,.
\]
Then $\frac{\epsi^2}{2}\Big(-i\nabla_X \,+ \,A_{\rm ext}(X)\Big)^2$ is self-adjoint on 
$W^2(\IR^n)$,  the second Sobolev space, since $-i\nabla_X$ is infinitesimally operator bounded
with respect to $-\Delta_X$. 
It follows that 
\begin{equation}\label{HEDEF}
H^\epsi = \frac{\epsi^2}{2}\Big(-i\nabla_X \,+ \,A_{\rm ext}(X)\Big)^2
 \otimes {\bf 1} + H_{\rm e}
\end{equation}
self-adjoint on $D(H^\epsi) = 
W^2(\IR^{n})\otimes\mathcal{H}_{\rm e} \cap D(H_{\rm e})$.

For $X\in\Lambda$, $\Lambda\subset\IR^n$ open, we require in addition some regularity
for $H_{\rm e}(X)$ as a function of $X$:
\begin{enumerate} \em
\item[\bf H$_k$\,\,\,] 
$H_{\rm e1}(\cdot)\in C^k_{\rm b} (\Lambda, \mathcal{L}(\mathcal{H}_{\rm e}))$.
\end{enumerate}

The exact value of $k$ will depend on whether $\Lambda = \IR^n$ or $\Lambda\subset \IR^n$.
For the type of Hamiltonian  considered in the introduction, cf.\ (\ref{SBO1}), all the above
conditions including  
Condition {\bf H$_k$} are easily checked and put constraints only on the smoothness of the
external potentials and on the smoothness and the decay of the charge distribution of the nuclei.
For point nuclei  {\bf H$_k$} fails and a suitable substitute would require a generalization
of the Hunziker distortion method of \cite{KMSW}.

We will be interested in subsets  of $\{(X,s)\in\Lambda\times\IR:s\in\sigma(H_{\rm e}(X)\}$ 
 which are isolated from the rest of the spectrum in the following sense.

\begin{enumerate} \em
\item[\bf S\,\,\,] 
For $X\in\Lambda$, let $\sigma_*(X)\subset \sigma(H_{\rm e}(X))$ be such that there are functions
$f_\pm\in C_{\rm b}(\Lambda,\IR)$ and a constant $d>0$ with 
\[
\big[ f_-(X) + d, f_+(X) -d \big] \cap \sigma_*(X) = \sigma_*(X)
\]
and 
\[
\big[ f_-(X) , f_+(X) \big] \cap \big(  \sigma(H_{\rm e}(X)\setminus\sigma_*(X)\big) = \emptyset\,.
\]
\end{enumerate}

We set 
$P_* = \int^\oplus_{\Lambda}\,dX\,P_*(X)$, where $P_*(X) = \1I_{\sigma_*(X)}(H_{\rm e}(X))$
is the spectral projection of $H_{\rm e}(X)$ with respect to  $\sigma_*(X)$.
As explained in the introduction we have to distinguish two cases.
\medskip

\noindent {\em (i) Globally isolated bands}
\smallskip

We assume $\Lambda = \IR^n$ and let  
\begin{equation} \label{HDIAGDEF}
H^\epsi_{\rm diag} := P_*\,H^\epsi\,P_*+ P_*^\perp\,H^\epsi\,P_*^\perp\,.
\end{equation}
Since we aim at a uniform result for the adiabatic theorem, we introduce
the  Sobolev spaces $W^{1,\epsi}(\IR^{n})$ and $W^{2,\epsi}(\IR^{n})$ 
with respect to the $\epsi$-scaled gradient, i.e.
\[
W^{1,\epsi}(\IR^{n}) := \left\{ \phi\in L^2(\IR^{n}):\|\phi\|_{W^{1,\epsi}} :=
\|\epsi\,|\nabla\phi|\,\| + \|\phi\|<\infty\right\}
\]
and
\[
W^{2,\epsi}(\IR^{n}) := \left\{ \phi\in L^2(\IR^{n}):\|\phi\|_{W^{2,\epsi}} :=
\|\epsi^2\Delta\phi\| + \|\phi\|<\infty\right\}\,.
\]
Alternatively we will project on finite total energies and define
$\mathcal{E}(H^\epsi) := \1I_{(-\infty,\mathcal{E}]}(H^\epsi)$ as the projection on total energies smaller
than $\mathcal{E}$.

\begin{mathe}{Theorem} \label{AdiabaticT} \em Assume  {\bf H$_3$} and  {\bf S} for $\Lambda=\IR^n$.
Then $H^\epsi_{\rm diag}$ is self-adjoint on the domain of $H^\epsi$. 
There are constants $C,\widetilde C<\infty$ such that for all $t\in\IR$
\begin{equation} \label{AdiabaticTE}
\left\| 
e^{-iH^\epsi {t/\epsi}} -
e^{-iH^\epsi_{\rm diag}{t/\epsi}}
\right\|_{\mathcal{L}(W^{2,\epsi} \otimes \mathcal{H}_{\rm e},   
\mathcal{H})} \leq \,\epsi\,C\,(1+|t|)^3
\end{equation}
 and for all $\mathcal{E}\in \IR$
\begin{equation} \label{AdiabaticTE2}
\left\| \left(
e^{-iH^\epsi {t/\epsi}} -
e^{-iH^\epsi_{\rm diag}{t/\epsi}}\right)\,\mathcal{E}(H^\epsi)\,
\right\|_{\mathcal{L}(   
\mathcal{H})} \leq \,\epsi\,\widetilde C\,(1+|\mathcal{E}|)\,(1+|t|)\,.
\end{equation}
$\mathcal{L}(W^{2,\epsi}\otimes \mathcal{H}_{\rm e},   
\mathcal{H})$ denotes the space of bounded linear operators from 
$W^{2,\epsi}\otimes \mathcal{H}_{\rm e}$ to $\mathcal{H}$ equipped with the
operator norm.
\end{mathe}

This result should be understood as  an adiabatic theorem 
for the subspaces Ran$P_*$ and Ran$P_*^\perp$, which are {\em not} spectral subspaces.
Let us point out  one immediate application of Theorem \ref{AdiabaticT}.
The behavior near band crossings is usually investigated using simplified models involving
only two energy bands and ignoring the rest of the spectrum,  
cf.\ \cite{HagedornCross,Gerard}.
Theorem \ref{AdiabaticT} shows that this strategy is indeed justified modulo errors of
order $\epsi$. 
\medskip

\noindent {\em (ii) Locally isolated bands}
\smallskip

 $\sigma_*(X)=E(X)$ is a nondegenerate eigenvalue for all $X\in\Lambda$.  
 $\Lambda$ may now be any open subset of $\IR^n$ and
for such a $\Lambda$ we assume {\bf H$_\infty$} and {\bf S}.
We also assume that $\Lambda$ is connected. Otherwise one could treat each connected
 component
separately.

It is easy to see that, given {\bf H$_\infty$} and {\bf S}, the family of projections 
$P_*(\cdot)\in C^\infty_{\rm b}(\Lambda,\mathcal{L}(\mathcal{H}_{\rm e}))$. 
However, in order to ``map'' the dynamics from Ran$P_*$ to $L^2(\Lambda)$
we need in addition a smooth version $\chi(\cdot) \in C^\infty_{\rm b}(\Lambda,\mathcal{H}_{\rm e})$
of  the normalized eigenvector  of $H_{\rm e}(X)$ with eigenvalue $E(X)$. In other words we require
the complex line bundle over $\Lambda$ defined by $P_*$ to be trivial.
This always holds for contractible $\Lambda$, but, as discussed below, 
 also for some relevant examples where $\Lambda$ is not contractible.

Given a smooth version of $\chi(X)$ with $\|\chi(X)\|=1$, 
one has Re$\langle \chi(X),\nabla_X \chi(X)\rangle = 0$, but, in general,
Im$\langle \chi(X),\nabla_X \chi(X)\rangle \not= 0$. 
In the following we distinguish two cases: Either it is possible
to achieve Im$\langle \widetilde \chi(X),\nabla_X \widetilde\chi(X)\rangle = 0$
by a smooth gauge transformation $\chi(X)\to \widetilde\chi(X) = e^{i\theta(X)}\chi(X)$
or not. In the latter case
\[
A_{\rm geo}(X) := -i\langle \chi(X),\nabla_X \chi(X)\rangle  
\]
is  the gauge potential of a  
connection on the trivial complex line bundle over $\Lambda$, the Berry connection,
and  has to be  taken  into account in the definition of the effective operator
\begin{equation}\label{HBO2}
H_{\rm BO}^\epsi := \frac{\epsi^2}{2}\Big( -i\nabla_X +  A_{\rm ext}(X) + 
 A_{\rm geo}(X)\Big)^2 + E(X)
\end{equation}
with domain $W^2(\IR^{n})$. Thus $A_{\rm geo}$  acts  as an additional external 
magnetic vector potential.
 Although  $A_{\rm ext}$ and $A_{\rm geo}$ appear in $H_{\rm BO}^\epsi$ with an
 $\epsi$ in front only, 
 and  therefore are not retained in the  semiclassical limit to leading order, they  do contribute
 to the solution of the Schr\"odinger equation for times of order $\epsi^{-1}$. 
If the full Hamiltonian is real in position representation, as it is the
case for the Hamiltonians considered in the introduction whenever $A_{\rm ext}=0$, 
then $\chi(X)$ can be chosen
 real-valued. If, in addition, $\Lambda$ is contractible, the existence of a smooth version 
of $\chi(X)$ with  Im$\langle \chi(X),\nabla_X\chi(X)\rangle = 0$
follows.

To define  $H_{\rm BO}^\epsi$ on $L^2(\IR^n)$ through (\ref{HBO2}), the functions 
$E(X)$ and $A_{\rm geo}(X)$, which are a priori defined
on $\Lambda$ only,  must be continued to functions on $\IR^n$.
Hence we arbitrarily extend $E(X)$ and $A_{\rm geo}(X)$ to functions in
$C^\infty_{\rm b}(\IR^n)$ by
 modifying them, if necessary, on $\Lambda\setminus(\Lambda-\delta/5)$ 
(cf.\  (\ref{GmaD})) for some $\delta>0$.
The parameter $\delta$ will be fixed in the formulation of Theorem \ref{MT} and will appear in
 several places. It controls how close the states are allowed to come to  $\partial\Lambda$.

The  generic example for the Berry phase  is a band crossings of
codimension 2 (cf.\ \cite{Berryphase,HagedornCross,Gerard}). 
If $E(X)$ is an isolated energy band  except for a  codimension 2 crossing, then 
$\Lambda = \IR^{n}\setminus\{$closed neighborhood of the crossing$\}$
is no longer contractible, but the line bundle is still trivial. 
Although the underlying Hamiltonian is real, the Berry connection cannot be gauged away.
Within the time-independent Born-Oppenheimer approximation
Herrin and Howland \cite{HerrinHowland} study
 a model with a nontrivial eigenvector bundle.

With the fixed  choice for  $\chi(X)$ we have
\begin{equation}\label{RanP}
 {\rm Ran}\, P_* = \left\{
\int_\Lambda^\oplus dX\, \phi(X)\chi(X);\, \phi\in L^2(\Lambda) \right\}\subset \mathcal{H}\,.
\end{equation}
Thus there is a natural identification $\mathcal U:$ Ran$P_*\to L^2(\IR^n)$
connecting the relevant subspace  on which the full quantum evolution takes place and
the Hilbert space $L^2(\IR^n)$  on which the effective Born-Oppenheimer evolution is
defined.
According to (\ref{RanP}), we set
\[
\mathcal{U}(\phi\chi) = \phi\,,\qquad{\rm i.e.}\quad (\,\mathcal{U}P_*\,\psi\,)(X) =
\langle\,\chi(X)\,,\,(P_*\,\psi)(X)\,\rangle_{\mathcal{H}_{\rm e}}\,.
\]
Its adjoint $\mathcal{U}^*: L^2(\IR^n)\to$ Ran$P_*$ is given by
\[
\mathcal{U}^*\phi = \int^\oplus_\Lambda\,dX\,\phi(X)\chi(X)\,.
\]
Clearly $\mathcal U$ is an isometry and $\mathcal{U}^*\mathcal{U}=\bf 1$
on Ran$P_*$. But  $\mathcal U$ is not surjective and thus not unitary.

By construction, $e^{-iH_{\rm BO}^\epsi{t/\epsi}}$ is a good approximation
to the true dynamics only as long as the wave function of the nuclei is  supported
in $\Lambda$ modulo 
errors of order  $\epsi$. Since $H_{\rm BO}^\epsi$ is a standard semiclassical operator,
the $X$-support of solutions of (\ref{BOSE}) 
 can be  calculated approximately from the  classical dynamics 
 generated by its principal symbol $H_{\rm cl} (q,p) = \frac{1}{2}p^2 +E(q)$
 on phase space $Z:=\IR^{n}\times\IR^{n}$,
\begin{equation}\label{classic}
\frac{d}{d t}q = p\,,\qquad \frac{d}{d t}p = -\nabla E(q)\,.
\end{equation}
 The solution flow to (\ref{classic}) 
exists for all times and will be denoted by $\Phi^t$.

In order to make these notions  more precise, we need to introduce some notation.
The Weyl quantization of $a\in C^\infty_{\rm b}(Z)$ is the linear operator
\[
\left( a^{\rm W,\epsi}\phi \right)(X) = (2\pi)^{-n} \int_{\IR^{n}}\,dY \,dk\,
a\left( \frac{X+Y}{2}, \epsi \,k\right) e^{-i(X-Y)\cdot k}\phi(Y)\,,
\]
 as acting on 
Schwartz functions. $a^{\rm W,\epsi}$ extends to $\mathcal{L}(L^2(\IR^{n}))$ with
 operator norm   bounded uniformly in $\epsi$ 
(cf., e.g., Theorem 7.11 in \cite{DS}).
The wave functions with phase space support in a compact  set $\Gamma\subset Z$ do not 
form a closed subspace of $L^2(\IR^{n})$. Hence we cannot project on this set.
In order to define 
approximate projections, let for
 $\Gamma\subset\IR^m$, $m\in\IN$, and  for $\alpha>0$ 
\begin{equation}\label{GmaD}
\Gamma-\alpha := \left\{ z\in \Gamma:\, \inf_{w\in \IR^m\setminus\Gamma}|w-z|\geq\alpha
\right\}\,.
\end{equation}

\begin{mathe}{Definition} \label{CDef} \em
An {\em approximate characteristic function} $\1I_{(\Gamma,\alpha)}\in C^\infty_{\rm b}(\IR^m)$ 
of a set $\Gamma\subset\IR^m$ with margin $\alpha$ is defined by the requirement that 
$\1I_{(\Gamma,\alpha)}|_{\Gamma-\alpha} =1$
and $\1I_{(\Gamma,\alpha)}|_{\IR^m\setminus\Gamma} =0$.

If $\1I_{(\Gamma,\alpha)}$ is an approximate characteristic function on phase space $Z$, then 
the corresponding {\em approximate projection} is defined as
its Weyl
quantization $\1I_{(\Gamma,\alpha)}^{\rm W,\epsi}$.
 We will say that
functions in Ran$\1I_{(\Gamma,\alpha)}^{\rm W,\epsi}$ have phase space support in 
$\Gamma$.
\end{mathe}

For $\Gamma\subset Z$ we will use the abbreviations
\begin{eqnarray*}
\Gamma_q& := &\left\{ q\in\IR^{n}: \,(q,p)\in\Gamma\,\,{\rm for\,some}\, p\in\IR^{n}\right\}\,,
\\
\Gamma_p& := &\left\{ p\in\IR^{n}: \,(q,p)\in\Gamma\,\,{\rm for\,some}\, q\in\IR^{n}\right\}\,.
\end{eqnarray*}

 Let the phase space support $\Gamma$ of the initial wave function   be  such that 
$\Gamma_q \subset \Lambda - \delta$.
Then the maximal time interval for which the $X$-support of the wave function of the nuclei
stays in $\Lambda$ up to errors of order $\epsi$ can be written as
\[
I_{\rm max}^\delta(\Gamma,\Lambda) := [T_-^\delta (\Gamma,\Lambda),
 T_+^\delta (\Gamma,\Lambda) ]\,,
\]
where the ``first hitting times'' $T_\pm$ are defined by the classical dynamics through
\[
T_+^\delta (\Gamma,\Lambda) :=
  \sup \Big\{t\geq 0:\, \big(\Phi^{s}(\Gamma)\big)_q  
\subseteq \Lambda-\delta\,\,\,
\forall\,s\in[0,t]\Big\}
\]
and $T_-^\delta (\Gamma,\Lambda)$ analogously for negative times. 
This are just the first  times for
a particle starting in $\Gamma$ to hit the boundary of $\Lambda-\delta$ when dragged along
the classical flow $\Phi^t$.

The following proposition, which is an immediate consequence of
Egorov's Theorem  \cite{RobertPaper,Robert}, shows that for times in 
$I_{\rm max}^\delta(\Gamma,\Lambda)$ the support of the wave function of the nuclei
stays indeed in $\Lambda-\delta$, up
to errors of order $\epsi$ uniformly on Ran$\1I_{(\Gamma,\alpha)}^{\rm W,\epsi}$
for any approximate projection $\1I_{(\Gamma,\alpha)}^{\rm W,\epsi}$.

\begin{mathe}{Proposition} \label{CLPROP} \em
Let $\Gamma\subset Z$ be such that $\Gamma_q\subset \Lambda-\delta$ and 
let $\1I_{\Lambda-\delta}$ denote multiplication with the characteristic
function of $\Lambda-\delta$ on $L^2(\IR^{n})$.
 For  any approximate projection $\1I_{(\Gamma,\alpha)}^{\rm W,\epsi}$
and any bounded interval
$I\subseteq I_{\rm max}^\delta(\Gamma,\Lambda)$
there is a constant $C<\infty$ such that for all $t\in I$
\[
\left\|
\big({\bf 1} - \1I_{\Lambda-\delta} \big)\, e^{-i H_{\rm BO}^\epsi {t/\epsi}}
\,\1I_{(\Gamma,\alpha)}^{\rm W,\epsi}
\right\|_{\mathcal{L}(L^2(\IR^n))} \leq\,C\,\epsi\,.
\]
\end{mathe}

An approximate projection on $\Gamma$ in 
$\mathcal{H}$ is defined as $P_\Gamma^\alpha :=
\mathcal{U}^*\,\1I_{(\Lambda,\delta)}\,\1I_{(\Gamma,\alpha)}^{\rm W,\epsi}\,\mathcal{U}\,P_*$, where 
$\1I_{(\Gamma,\alpha)}^{\rm W,\epsi}$ is an approximate 
projection on $\Gamma$ according to Definition \ref{CDef}  and 
$\1I_{(\Lambda,\delta)}$ is an approximate characteristic function for $\Lambda$.
Using the latter instead of the sharp cutoff  from $\mathcal U^*$ makes Ran$P_\Gamma^\alpha$
a bounded set in $W^{2,\epsi}\otimes \mathcal{H}_{\rm e}$ whenever $\Gamma_p$ is a bounded set.

\begin{mathe}{Theorem} \label{MT} \em
Assume  {\bf H$_\infty$} and  {\bf S} with dim(Ran$P_*(X))=1$ for some open  
$\Lambda\subseteq\IR^n$.
Let $\Gamma\subset Z$ be such that $\Gamma_q\subset\Lambda-\delta$ for some $\delta>0$
and $\Gamma_p$ bounded.
For any approximate projection $P_\Gamma^\alpha$
and any bounded interval $I\subseteq I_{\rm max}^\delta (\Gamma,\Lambda)$ 
 there is a constant $C<\infty$ such that for all $t\in I$
\begin{equation} \label{MTE}
\left\| \left( e^{-iH^\epsi{t/\epsi}} - \mathcal{U}^*\,e^{-iH^\epsi_{\rm BO}{t/\epsi}}
\,\mathcal{U}
\right)\, P_\Gamma^\alpha \right\|_{\mathcal{L}(\mathcal{H})}\leq C\epsi\,.
\end{equation}
\end{mathe}

Theorem \ref{MT} establishes
 that the electrons adiabatically follow the motion of
the nuclei up to errors of order $\epsi$ as long as the leaking through the boundary of $\Lambda$
is small. The semiclassics was used  only  to control such a leaking uniformly.
However, for $H^\epsi_{\rm BO}$
the limit $\epsi\to 0$  is a semiclassical limit and, as discussed in the following section, 
beyond the mere support of the wave function more  detailed information  is available.

\section{Semiclassics for a single band}

The semiclassical limit of Equation (\ref{BOSE}) with a Hamiltonian of the form
(\ref{HBO2}) is well understood and there is a variety of different approaches. 
For example one can  construct approximate solutions $\phi_{q(t)}$
of (\ref{BOSE}) which are localized along a classical trajectory $q(t)$, i.e.\ along a solution
of (\ref{classic}). Then it follows from  Theorem \ref{MT} that
$\phi_{q(t)}\chi$ is a solution of the full Schr\"odinger equation, 
(\ref{SE2}), up to an error of order $\epsi$ 
as long as $q(t)\in \Lambda-\delta$. Roughly speaking,
this coincides with the result of  Hagedorn \cite{Hagedorn80}.
In applications the assumption that
the wave function of the nuclei is well described by a coherent state seems to be rather restrictive
and a more general approach to the semiclassical analysis of a Schr\"odinger equation of the form
(\ref{BOSE}) is to consider the distributions of semiclassical  observables, i.e.\
of operators obtained as Weyl quantization $a^{\rm W,\epsi}$ of classical phase space functions 
$a:Z\to \IR$.

 Consider a general initial wave function $\phi^\epsi\in L^2(\IR^n)$, such that
$\phi^\epsi$ corresponds to a probability measure $\rho_{\rm cl}(dq\,dp)$ 
on phase space in the  sense that
for all semiclassical observables with symbols $a\in C^\infty_{\rm b}(Z)$ 
\begin{equation} \label{distdef}
\lim_{\epsi\to 0} \left|
\langle \phi^\epsi, \,a^{\rm W,\epsi}\,\phi^\epsi \rangle -
\int_{Z}\,a(q,p)\,\rho_{\rm cl}(dq\,dp)\,\right| = 0\,.
\end{equation}
The definition is equivalent to saying that the Wigner transform of $\phi^\epsi$ 
converges to $\rho_{\rm cl}$
weakly on test functions in $C^\infty_{\rm b}(Z)$ \cite{LionsPaul}.
 An immediate application of Egorov's theorem yields
\begin{equation} \label{distr}
\lim_{\epsi\to 0} \left|
\langle \phi^\epsi,\, e^{iH^\epsi_{\rm BO}{t/\epsi}}\,a^{\rm W,\epsi}\,
 e^{-iH^\epsi_{\rm BO}{t/\epsi}}
\,\phi^\epsi \rangle -
\int_{Z}\,(a\circ\Phi^t)(q,p)\,\rho_{\rm cl}(dq\,dp)\,\right| = 0
\end{equation}
uniformly on bounded intervals in time, where we recall that $\Phi^t$ is the flow 
generated by (\ref{classic}). In (\ref{distr})
one can of course shift the time evolution from the observables to the states 
on both sides and write instead
\begin{equation}\label{distt}
\lim_{\epsi\to 0} \left|
\langle \phi^\epsi_t, \,a^{\rm W,\epsi}\,\phi^\epsi_t \rangle -
\int_{Z}\,a(q,p)\,\rho_{\rm cl}(dq\,dp,t)\,\right| = 0\,.
\end{equation}
Here $\phi^\epsi_t= e^{-iH_{\rm BO}{t/\epsi}}\phi^\epsi$ and $\rho_{\rm cl}(dq\,dp,t) = \big(
\rho_{\rm cl} \circ \Phi^{-t} \big)(dq\,dp)$ is the initial distribution $\rho_{\rm cl}(dq\,dp)$
transported along the classical flow.
Thus with respect to certain type of experiments the system described by the wave function
$\phi^\epsi_t$ behaves like a classical system.

For a molecular system the object of real interest is the left hand side of (\ref{distt}) with
$\phi^\epsi_t$ replaced by the solution $\psi_t^\epsi$ of the full Schr\"odinger equation
and $a^{\rm W,\epsi}=:a_{\rm BO}^\epsi$ as acting on $L^2(\IR^n)$ replaced by 
$a^{\rm W,\epsi} \otimes {\bf 1}$ as acting on
$\mathcal{H}$.
In order to compare the  expectations of  $a_{\rm BO}^\epsi$ with 
the expectations 
of $a^{\rm W,\epsi} \otimes {\bf 1} $, we need  the following proposition.

\begin{mathe}{Proposition} \label{MT2} \em
In addition to the assumptions of Theorem \ref{MT} let
 $a\in C^\infty_{\rm b}(Z)$ with
\begin{equation} \label{F2Norm}
 \int\,d\xi\, \sup_{x\in \IR^n}\,|\xi|\,| 
\widehat a^{(2)} (x,\xi)|\,<\infty\,,
\end{equation}
where\, $\widehat {\,\,\,\,}^{(2)}$\, denotes Fourier transformation in the second argument.
Then there is a constant $C<\infty$ such that 
\[
\left\| \left( a^{\rm W,\epsi}\otimes {\bf 1}\, - \,\mathcal{U}^*\,a^{\rm W,\epsi}\,\mathcal{U}
\right)\,\1I_{\Lambda-\delta}P_*\,\right\|\,\leq\,C \,\epsi\,.
\]
\end{mathe}

For the proof of Proposition \ref{MT2} see the end of Section 4.2.
With its help we obtain the semiclassical limit for the nuclei as governed by the full Hamiltonian.

\begin{mathe}{Corollary} \label{Cor1}\em
Let $\Gamma$ and $I$ be as in Theorem \ref{MT}.
Let $\psi^\epsi\in\mathcal H$ such that (\ref{distdef}) is satisfied for $\phi^\epsi :=
\mathcal{U}P_*\psi^\epsi$  for  some
$\rho_{\rm cl}$ with supp$\rho_{\rm cl}\subset\Gamma-\alpha$.
Let $\psi^\epsi_t =
e^{-iH^\epsi{t/\epsi}}\psi^\epsi$ then  for all $a\in C^\infty_{\rm b}(Z)$ which
satisfy (\ref{F2Norm}) 
\begin{equation} \label{CorEq}
\lim_{\epsi\to 0} \left|
\langle \psi^\epsi_t, \,(a^{\rm W,\epsi}\otimes{\bf 1})\,\psi^\epsi_t \rangle -
\int_{Z}\,a(q,p)\,\rho_{\rm cl}(dq\,dp,t)\,\right| = 0
\end{equation}
uniformly for $t\in I$.
\end{mathe}

Translated to the language of Wigner measures Corollary \ref{Cor1} states the following. 
Let us define the marginal Wigner transform for the nuclei as
\[
W_{\rm nuc}^\epsi(\psi^\epsi_t)(q,p) :=  (2\pi)^{-n} \int_{\IR^{n}}\,dX\,e^{iX\cdot p}
\,\langle  \psi^{\epsi *}_t (q+\epsi X/2) , 
\psi^{\epsi}_t (q-\epsi X/2)\rangle_{\mathcal{H}_{\rm e}}\,.
\]
Then, whenever $W_{\rm nuc}^\epsi(P_*\psi^\epsi_0)(q,p)\,dq\,dp$ converges weakly to some 
probability  measure  $\rho_{\rm cl}(dq\,dp)$,  $W_{\rm nuc}^\epsi(P_*\psi^\epsi_t)(q,p)\,dq\,dp$
converges weakly to $(\rho_{\rm cl} \circ \Phi^{-t} \big)(dq\,dp)$.

Corollary \ref{Cor1} follows by applying first 
 Proposition \ref{MT2} and then Theorem \ref{MT} to the left hand side in the difference
(\ref{CorEq}), where we note that
$\lim_{\epsi\to 0}\|(1-P_{\Gamma}^\alpha)\psi^\epsi\|=0$ and thus also 
$\lim_{\epsi\to 0}\|(1-P_{\Lambda-\delta'})\psi^\epsi_t\|=0$ for any $\delta'<\delta$. This
 yields the left hand side of (\ref{distr}) and thus (\ref{CorEq}). 

We mention some standard examples of initial wave functions
$\phi^\epsi$ of the nuclei which approximate  certain classical distributions.
The initial wave function for the full system is, as before, recovered as $\psi^\epsi 
= \mathcal{U}^* \phi^\epsi = \phi^\epsi(X)\chi(X)$.
In these examples one  regains some control
on the rate of convergence with respect to $\epsi$ which  was lost in (\ref{distdef}). 
\medskip

\noindent
{\em (i) Wave packets tracking a classical trajectory.}
\smallskip

For $\phi\in L^2(\IR^n)$ let
\[
\phi^\epsi_{q_0,p_0} (X) = \epsi^{-\frac{n}{4}}\,e^{-i\frac{p_0\cdot(X-q_0)}{\epsi}}
\phi\big(\frac{X-q_0}{\sqrt{\epsi}}\big)\,.
\]
Then $|\phi^\epsi_{q_0,p_0} (X)|^2$ is sharply peaked at $q_0$ for $\epsi$ small
and its $\epsi$-scaled Fourier transform is sharply peaked at $p_0$. Thus one expects
that the corresponding classical distribution is given by $\delta(q-q_0)\delta(p-p_0)\,dq\,dp$.
As was shown, e.g.\ in \cite{SpohnT}, this is indeed true for $\phi\in L^2(\IR^n)$
such that $\phi,|x|\phi,\widehat\phi,|p|\widehat\phi\in L^1(\IR^n)$.
Then Corollary \ref{Cor1} holds with  (\ref{CorEq})   replaced by
\begin{eqnarray} \lefteqn{
 \left|
\langle \psi^\epsi_t, \,(a^{\rm W,\epsi}\otimes{\bf 1})\,\psi^\epsi_t \rangle -
a(q(t),p(t))
\,\right|}\nonumber\\ &= &
 O(\sqrt{\epsi})\,\left(\|\phi\|^2_{L^2}\, + \, \|\phi\|_{L^1}\,\||p|\widehat\phi\|_{L^1}\,+\,
\||x|\phi\|_{L^1}\,\|\widehat\phi\|_{L^1}\right)
\,,\label{iEx}
\end{eqnarray}
where $(q(t),p(t))$ is the solution of the classical dynamics with initial condition $(q_0,p_0)$.
(\ref{iEx}) generalizes Hagedorn's first order result in \cite{Hagedorn80} to a larger class
of  localized wave functions.
\medskip

\noindent
{\em (ii) Either sharp momentum or sharp position.}
\smallskip

For $\phi\in L^2(\IR^n)$ let
\[
\widehat \phi^\epsi_{p_0}(p) = \widehat \phi(p-p_0/\epsi)\,,
\]
where \,$\widehat{ }$\, denotes the $\epsi$-scaled Fourier transformation, 
then the corresponding classical distribution is $\rho_{\rm cl}(dq\,dp) = 
\delta(p-p_0)|\phi(q)|^2\,dq\,dp$. Note that the absolute value of $\phi$ does not depend on
$\epsi$ in that case.
Equivalently one defines
\[
\phi^\epsi_{q_0}(X) =  \epsi^{-\frac{n}{2}}
\phi\big(\frac{X-q_0}{\epsi}\big)
\]
and obtains $\rho_{\rm cl}(dq\,dp) = \delta(q-q_0)|\widehat\phi(p)|^2\,dq\,dp$.
In both cases one finds that the difference in  (\ref{CorEq}) 
is bounded  a constant times either
$\epsi \big(\|\phi\|_{L^2}^2 + \|\phi\|_{L^1}\||p|\widehat\phi\|_{L^1}\big)$
for $\phi^\epsi_{p_0}$ or 
$\epsi \big(\|\phi\|_{L^2}^2 + \||x|\phi\|_{L^1}\|\widehat\phi\|_{L^1}\big)$
 for $\phi^\epsi_{q_0}$. 
\medskip

\noindent {\em (iii) WKB wave functions.}
\smallskip

For $f\in L^2(\IR^n)$ and $S\in C^1(\IR^n)$  both real valued let
\[
\phi^\epsi(X) = f(X)\,e^{i\frac{S(X)}{\epsi}}\,,
\]
then $\rho_{\rm cl}(dq\,dp)= f^2(q)\,\delta(p-\nabla S(q))\,dq\,dp$.
In this case one expects that
(\ref{CorEq})  is bounded as $\sqrt{\epsi}$, which has been shown in 
\cite{SpohnT} for a smaller set of test functions.

\section{Proofs}

\subsection{Globally isolated bands}

We collect some immediate consequences of  {\bf H$_3$} and {\bf S}.
Using the Riesz formula
\begin{equation}\label{riesz}
P_*(X) = -\frac{1}{2\pi i} \oint_{\gamma(X)}\,d\lambda\,R_\lambda(H_{\rm e}(X))\,,
\end{equation}
with $\gamma(X)$ a smooth curve in the complex plain circling $\sigma_*(X)$ only
and $R_\lambda(H_{\rm e}(X)) = (H_{\rm e}(X)-\lambda)^{-1}$,
one easily shows that
$P_*(\cdot)\in C^2_{\rm b}(\IR^{n},\mathcal{L}(\mathcal{H_{\rm e}}))$. 
Assumption {\bf S} enters at this point, since it allows to chose $\gamma(X)$ locally independent
of $X$. Hence, when taking  derivatives with respect to $X$ in (\ref{riesz}), one only needs
to differentiate the integrand. In particular one finds that
\begin{eqnarray} \displaystyle \lefteqn{
P_*^\perp(X)(\nabla_XP_*)(X)P_*(X) =}\nonumber \\ \displaystyle
&\displaystyle \frac{1}{2\pi i} \oint_{\gamma(X)}\,d\lambda\,
R_\lambda(H_{\rm e}(X))\,P_*^\perp(X)\,(\nabla_XH_{\rm e})(X)\, R_\lambda(H_{\rm e}(X))\,P_*(X)\,.
\label{KI1}
\end{eqnarray}
Since 
 $P_*(X)(\nabla_XP_*)(X)P_*(X)= P^\perp_*(X)(\nabla_XP_*)(X)P^\perp_*(X) =  0$, 
which follows from 
$(\nabla_X P_*)(X)=(\nabla_X P_*^2)(X) = (\nabla_XP_*)(X)P_*(X) +P_*(X)  (\nabla_XP_*)(X)$,
we have that
\begin{equation}
(\nabla_XP_*)(X)
= P_*^\perp(X)(\nabla_XP_*)(X)P_*(X) +  \,\,{\rm adjoint}
\,.\label{G1}
\end{equation}
In (\ref{G1}) and in the following ``$+$ adjoint'' means that the adjoint operator of the first
term in a sum is added.

Starting   with  (\ref{AdiabaticTE}), we find, at the moment formally,   that
\begin{eqnarray} 
\displaystyle \lefteqn{ \displaystyle \hspace{-1cm}
e^{-iH^\epsi_{\rm diag} {t/\epsi}} -
e^{-iH^\epsi{t/\epsi}}\, =\, e^{-iH^\epsi_{\rm diag} {t/\epsi}}
\left({\bf 1}- e^{iH^\epsi_{\rm diag} {t/\epsi}}\,e^{-iH^\epsi{t/\epsi}}
\right)
 = } \nonumber\\\displaystyle
&\displaystyle = &\displaystyle i\,  e^{-iH^\epsi_{\rm diag} {t/\epsi}}\,\int_0^{t/\epsi}\,ds\,
 e^{iH^\epsi_{\rm diag} s}\, \big( H^\epsi - H^\epsi_{\rm  diag} \big) e^{-iH^\epsi s}\,,
\label{I1}
\end{eqnarray}
where 
\begin{eqnarray}
 H^\epsi - H^\epsi_{\rm  diag}
& =& P_*^\perp\,H^\epsi\,P_* + \,\,{\rm adjoint}
\nonumber\\
& = & P_*^\perp \left[\frac{\epsi^2}{2}\Big(-i\nabla_X + A_{\rm ext}(X)\Big)^2, P_*\right] P_* + \,\,{\rm adjoint}\,.\label{HD1}
\end{eqnarray}
Let $D_A := -i\nabla_X + A_{\rm ext}(X)$.
Then the commutator is easily calculated as 
\begin{eqnarray}\displaystyle
 \left[\frac{\epsi^2}{2}(D_A\otimes {\bf 1})^2, P_*\right]& =&- i\,\epsi \,
 (\nabla_X P_*)\,\cdot\,(\epsi D_A \otimes{\bf 1}) + O(\epsi^2)\label{K0}\\
& =&-  \epsi \,
 (\nabla_X P_*)\,\cdot\,(\epsi\nabla_X\otimes{\bf 1}) + O(\epsi^2)\,, \label{K1}
\end{eqnarray}
where $O(\epsi^2)$ holds in the norm of $\mathcal{L(H,H)}$ as $\epsi\to 0$. 
For (\ref{K0}) and (\ref{K1}) it was used that $A_{\rm ext}(X)$ and $P_*(X)$ are both 
differentiable with bounded derivatives and that $A_{\rm ext}(X)$ commutes with $P_*$.

Before we can continue, we need to justify (\ref{I1}) by showing that $H_{\rm diag}^\epsi$
is self-adjoint on $D(H^\epsi)$. To see this, note that $-i\epsi\nabla_X$ is 
bounded with respect to $\epsi^2\Delta_X$ with relative bound $0$ and that
for $\psi\in D(H^\epsi)$
\begin{eqnarray}
\| (\epsi^2\Delta_X\otimes{\bf 1})\,\psi\|
&\leq &\,c_1\,\left(
\| (\epsi^2 D_A^2 \otimes{\bf 1}) \,\psi\| +\|\psi\|\right) \nonumber\\
&\leq &\,c_2\, \left(
\| ( \epsi^2 D_A^2 \otimes{\bf 1} + {\bf 1}\otimes H_0 ) \,\psi\| 
+\|\psi\| \right)\nonumber\\
& \leq & \,c_3\,
\left( \| H^\epsi\,\psi\| +\|\psi\|\right)\,,\label{HDD}
\end{eqnarray}
where we used that $H_{\rm e0}$ is bounded from below and that $H_{\rm e1}$ is bounded.
Hence $H^\epsi-H^\epsi_{\rm diag}$ is infinitesimally operator bounded with respect to
$H^\epsi$, consequently $H^\epsi_{\rm diag}$ is  self-adjoint on $D(H^\epsi)$ and thus
(\ref{I1}) holds on $D(H^\epsi)$.

(\ref{HD1}) and (\ref{K1}) in (\ref{I1}) give
\begin{eqnarray} \label{I2}
\displaystyle \lefteqn{ \displaystyle P_*^\perp
\left( e^{-iH^\epsi_{\rm diag} {t/\epsi}} -
e^{-iH^\epsi{t/\epsi}}\right)\, = } \\\displaystyle
&\displaystyle = &\displaystyle - i \epsi\,
 e^{-iH^\epsi_{\rm diag} {t/\epsi}}\,\int_0^{t/\epsi}\,ds\,
 e^{iH^\epsi_{\rm diag} s}\, P_*^\perp\,
 (\nabla_X P_*)\,P_*\,\cdot\,(\epsi\nabla_X\otimes{\bf 1})
 \, e^{-iH^\epsi s}\,+O(\epsi)|t| \nonumber\,,
\end{eqnarray}
where we used that
 the term of order $O(\epsi^2)$ in (\ref{K1}) yields a term
of order $O(\epsi)|t|$ after integration, since all other expressions in the integrand
are bounded uniformly in time
and the domain of integration grows like $t/\epsi$. In (\ref{I2}) and in the following
we omit the adjoint term from (\ref{HD1}) and thus consider the difference of the groups
projected on Ran$P_*^\perp$ only. The argument for the difference projected 
on Ran$P_*$ goes through analogously by taking adjoints at the appropriate places.

Now $\epsi(\nabla_X P_*)\,\cdot\,(\epsi\nabla_X\otimes{\bf 1})$ is only $O(\epsi)$
in the norm of  $\mathcal{L}(W^{1,\epsi}\otimes \mathcal{H}_{\rm e},\mathcal{H})$ and
thus, according to the naive argument, only  $O(1)|t|$ after integration.
As in \cite{HST} and \cite{SpohnT} we proceed by writing
$ (\nabla_X P_*)\,\cdot\,(\epsi\nabla_X\otimes{\bf 1})$ as the  
commutator of a bounded operator 
$B$  with $H^\epsi$ modulo
terms of order $O(\epsi)$. This is in analogy to the proof of the time-adiabatic theorem
\cite{Kato} and allows one to write the first order part of the
integrand in (\ref{I2}) as the time derivative of a bounded operator and, as a consequence, to
do the integration without losing one order in $\epsi$.
 
In view of (\ref{KI1}) we define
\begin{equation} \label{BtildeDef}
\widetilde B(X) :=   \frac{1}{2\pi i} \oint_{\gamma(X)}\,d\lambda\,
R_\lambda(H_{\rm e}(X))^2\,P_*^\perp(X)\,(\nabla_X H_{\rm e})(X)\, R_\lambda(H_{\rm e}(X))\,P_*(X)\,.
\end{equation}
An easy calculation shows that
\begin{equation} \label{K2}
\left[ H_{\rm e}, \widetilde B \right] = -\,P^\perp_*\, (\nabla_X P_*)\,P_*\,.
\end{equation}
By assumption $\partial_{X_j} H_{\rm e}(X)\in C^2(\IR^n,\mathcal{L}(\mathcal{H}_{\rm e}))$, 
$j=1,\ldots, n$, hence $\widetilde B_j(X)\in C^2(\IR^n,\mathcal{L}(\mathcal{H}_{\rm e}))$ 
 and thus
\begin{equation}\label{K3}
\left[ \frac{\epsi^2}{2}D_A^2 \otimes{\bf 1}, \widetilde B \right] =
- \epsi\,(\nabla_X\widetilde B)\,\cdot \,(\epsi\nabla_X\otimes{\bf 1}) +O(\epsi^2)
 = O(\epsi)
\end{equation}
in the norm of $\mathcal{L}(W^{1,\epsi}\otimes \mathcal{H}_{\rm e},\mathcal{H})$.
(\ref{K2}) and (\ref{K3}) combined yield that
\[
\left[ H^\epsi,\widetilde B\right] =  -\,P^\perp_*\, (\nabla_X P_*)\,P_* + O(\epsi)
\]
with $O(\epsi)$ in the norm of $\mathcal{L}(W^{1,\epsi}\otimes \mathcal{H}_{\rm e},\mathcal{H})$.
Since $\nabla_X H_{\rm e}\in \mathcal{L(H)}$,  a short calculation shows that
$[H^\epsi,\epsi\nabla_X\otimes {\bf 1}] =   O(\epsi)$ in 
$\mathcal{L}(W^{1,\epsi}\otimes \mathcal{H}_{\rm e},\mathcal{H})$.
Hence we  define
\[
B: = \widetilde B\,\cdot\, (\epsi\nabla_X\otimes{\bf 1})
\]
and obtain
\[
\left[ H^\epsi, B\right] =  -\,P^\perp_*\, (\nabla_X P_*)\,P_*\,\cdot\,  
(\epsi\nabla_X\otimes{\bf 1})  + O(\epsi)
\]
with $O(\epsi)$ in the norm of $\mathcal{L}(W^{1,\epsi}\otimes \mathcal{H}_{\rm e},\mathcal{H})$.
Let
\[
B(s) = e^{iH^\epsi s}\,B\,e^{-iH^\epsi s}
\]
then 
\[
-i\frac{d}{ds}\,B(s) =  e^{iH^\epsi s}\,[H^\epsi,B]\,e^{-iH^\epsi s}\,.
\]

Continuing  (\ref{I2}), we have
\begin{eqnarray} \label{I25}
\displaystyle \lefteqn{ \displaystyle P_*^\perp\,
\left( e^{-iH^\epsi_{\rm diag} {t/\epsi}} -
e^{-iH^\epsi{t/\epsi}}\right)\, = } \nonumber
\\\displaystyle
&\displaystyle = &\displaystyle  i\, \epsi\,
 e^{-iH^\epsi_{\rm diag} {t/\epsi}}\,\int_0^{t/\epsi}\,ds\,
e^{iH^\epsi_{\rm diag} s}\, \left[ H^\epsi, B\right]\, e^{-iH^\epsi s}\,+O(\epsi)(|t|+|t|^2)
\nonumber\\\displaystyle
&\displaystyle = &\displaystyle \epsi\,
 e^{-iH^\epsi_{\rm diag} {t/\epsi}}\,\int_0^{t/\epsi}\,ds\,
e^{iH^\epsi_{\rm diag} s}\, e^{-iH^\epsi s}\,\left( \frac{d}{ds}\,B(s)\right)\,+O(\epsi)(|t|+|t|^2)
\,,
\end{eqnarray} 
where  $O(\epsi)$ holds now in the norm of 
$\mathcal{L}(W^{1,\epsi}\otimes \mathcal{H}_{\rm e},\mathcal{H})$. The additional
factor of $|t|$ in (\ref{I25}) comes from the fact that 
\begin{equation} \label{I26}
\left\|e^{-iH^\epsi s}\right\|_{\mathcal{L}(W^{1,\epsi}\otimes \mathcal{H}_{\rm e})}
\leq \,c\,(1+\epsi\,|s|)
\end{equation}
for some constant $c<\infty$, i.e.\ the scaled momentum of the nuclei may grow in time.
Using $\|A_{\rm ext}\|_\infty=C<\infty$ and  
\[
\left\| \left[ (\epsi D_A\otimes{\bf 1} ), H^\epsi\, \right]\right\|_{\mathcal{L(H)}}\,\leq\, 
\widetilde C \,\epsi\,,
\]
(\ref{I26}) follows from
\begin{eqnarray*}\lefteqn{
\hspace{-2cm}
\left\| (-i\epsi\nabla_X\otimes{\bf 1}) \,e^{-iH^\epsi s}\, \psi \right\|
\leq 
\left\|(\epsi D_A\otimes{\bf 1} )  \,e^{-iH^\epsi s}\, \psi \right\|
+ \left\|(\epsi A_{\rm ext}\otimes{\bf 1} )  \,e^{-iH^\epsi s}\, \psi \right\|}
\\
&\leq&\,
\left\|(\epsi D_A\otimes{\bf 1} )  \, \psi \right\| + 
\left\| \left[ (\epsi D_A\otimes{\bf 1} ), \,e^{-iH^\epsi s}\right]\, \psi \right\|
+ C \left\|\psi\right\|
\\
&\leq &\,
\left\| (-i\epsi\nabla_X\otimes{\bf 1})\, \psi \right\|\, +\, \widetilde C \,\epsi\,|s| \,\|\psi\|\,
+\,2\,C\,\|\psi\|
\end{eqnarray*}
for $\psi\in W^{1}\otimes \mathcal{H}_{\rm e}$.

Finally, continuing (\ref{I25}), integration by parts yields
\begin{eqnarray} \label{I3}
\displaystyle \lefteqn{ \displaystyle P^\perp_*\,
\left( e^{-iH^\epsi_{\rm diag} {t/\epsi}} -
e^{-iH^\epsi{t/\epsi}}\right) = } \nonumber\\\displaystyle
&\displaystyle = &\displaystyle  \epsi\,
 e^{-iH^\epsi_{\rm diag} {t/\epsi}}\,\int_0^{t/\epsi}\,ds\,
e^{iH^\epsi_{\rm diag} s}\, e^{-iH^\epsi s}\,\left( \frac{d}{ds}\,B(s)\right)\, +O(\epsi)
(|t|+|t|^2) \nonumber\\
&\displaystyle = &\displaystyle 
\epsi\,\left(B\, e^{-iH^\epsi{t/\epsi}}
-  e^{-iH^\epsi_{\rm diag} {t/\epsi}}\,B\right) \nonumber\\
&&+ \,i\,\epsi\, e^{-iH^\epsi_{\rm diag} {t/\epsi}}\,\int_0^{t/\epsi}\,ds\,
 e^{iH^\epsi_{\rm diag} s}\,\big(H^\epsi-H^\epsi_{\rm diag}\big)
 B\,e^{-iH^\epsi s}\,+O(\epsi)(|t|+|t|^2) \nonumber\\
&=& O(\epsi)\big(1+|t|\big)^3\,,
\end{eqnarray}
where $O(\epsi)$ holds in the norm of 
$\mathcal{L}(W^{2,\epsi}\otimes \mathcal{H}_{\rm e},\mathcal{H})$. For the
last equality we used that $B$ is bounded in 
$\mathcal{L}(W^{2,\epsi}\otimes \mathcal{H}_{\rm e}, \mathcal{H})$ as well as in 
$\mathcal{L}(W^{2,\epsi}\otimes \mathcal{H}_{\rm e},W^{1,\epsi}\otimes \mathcal{H}_{\rm e})$ 
uniformly with respect to $\epsi$,  $H^\epsi-H^\epsi_{\rm diag}$ is $O(\epsi)$
in $\mathcal{L}(W^{1,\epsi}\otimes \mathcal{H}_{\rm e},\mathcal{H})$, 
 as we saw in (\ref{HD1}) and (\ref{K1}), and 
\begin{equation} \label{I27}
\left\|e^{-iH^\epsi s}\right\|_{\mathcal{L}(W^{2,\epsi}\otimes \mathcal{H}_{\rm e})}
\leq \,c\,(1+\epsi\,|s|)^2
\end{equation}
for some constant $c<\infty$. (\ref{I27}) follows from arguments similar to those
used in the proof of (\ref{I26}). 

We are left to prove (\ref{AdiabaticTE2}). This follows from exactly the same proof using
that $\mathcal{E}(H^\epsi)$ commutes with $e^{-iH^\epsi s}$ and that, according to (\ref{HDD}),
\[
\| (\epsi^2\Delta_X\otimes{\bf 1})\,\mathcal{E}(H^\epsi)\,\psi\|
\leq  \,c_3\,
\left( \| H^\epsi\,\mathcal{E}(H^\epsi)\,\psi\| +\|\psi\|\right)
\leq\, c_4 \,(|\mathcal{E}|+1)\,\|\psi\|\,.
\]

\subsection{Locally isolated bands}

To prove Theorem \ref{MT} we proceed along the same lines as in the previous section, with the one
modification that we use Proposition \ref{CLPROP} to control the flux out of $\partial\Lambda$.
However, one cannot use $P_* = \int^\oplus_\Lambda\,dX\, P_*(X)$ to define $H^\epsi_{\rm diag}$ 
anymore, because the functions in its range
would not be in the range of $H^\epsi$ and some smoothing in the cutoff is needed. 
For $i\in\{0,1,2,3\}$ let $\1I_i=\1I_{(\Lambda-\frac{4-i}{5}\delta, \frac{1}{5}\delta)}$ be approximate
characteristic functions according to Definition \ref{CDef}.
 Then the smoothed projections are defined with
$P_i(X) = \1I_i(X)\,P_*(X)$ as 
$P_i = \int^\oplus\,dX\, P_i(X)$. 
In the following it will be used that for $i<j$ we have $P_iP_j=P_jP_i=P_i$ and hence
$(1-P_j)P_i =P_i(1-P_j) =0$.

Proposition \ref{CLPROP} yields
\begin{equation}\label{Neu1}
\left( e^{-iH^\epsi{t/\epsi}} - \mathcal{U}^*\,e^{-iH^\epsi_{\rm BO}{t/\epsi}}
\,\mathcal{U}
\right)\, P_\Gamma^\alpha \,=\,
\left( e^{-iH^\epsi{t/\epsi}} - P_1 \,\mathcal{U}^*\,e^{-iH^\epsi_{\rm BO}{t/\epsi}}
\,\mathcal{U}
\right)\, P_\Gamma^\alpha + O(\epsi)\,.
\end{equation}

We make also use of the fact
 that the phase space support of the initial wave function lies in $\Gamma$ and has thus 
bounded energy with respect to $H_{\rm cl}$. 
Let $E:= \sup_{z\in\Gamma}H_{\rm cl}(z)<\infty$, let $\1I_{((-\infty,E+\alpha),\alpha)}$
be a smooth characteristic function on $\IR$ and let 
$\mathcal{E}:= \Big(\1I_{((-\infty,E+\alpha),\alpha)}(H_{\rm cl}(\cdot))\Big)^{\rm W,\epsi}$.
Then standard results from semiclassical analysis imply the following relations.

\begin{mathe}{Proposition}\em \label{PNeu}
\begin{enumerate}
\item[(a)] $\1I_{(\Gamma,\alpha)}^{\rm W,\epsi} = 
\mathcal{E}\,\1I_{(\Gamma,\alpha)}^{\rm W,\epsi}+O(\epsi)$;
\item[(b)] $e^{-iH^\epsi_{\rm BO}{t/\epsi}} \,\mathcal{E} = \mathcal{E}\,
e^{-iH^\epsi_{\rm BO}{t/\epsi}} + O(\epsi)$ uniformly for $t\in I$;
\item[(c)] $[\,H_{\rm BO}^\epsi,\, \mathcal{E}\,] = O(\epsi^2)$;
\item[(d)] $\mathcal{E}\in\mathcal{L}(L^2(\IR^n),W^{2,\epsi})$. 
\end{enumerate}
In (a)--(c) $O(\epsi)$ resp.\ $O(\epsi^2)$ hold in the norm of  $\mathcal{L}(L^2(\IR^n))$.
\end{mathe}
Proposition (\ref{PNeu}) (a), (c) and (d) are direct consequences of the  product rule for
pseudo-differential operators (see, e.g., \cite{Robert,DS}) and (b) is again Egorov's Theorem.

Using Proposition \ref{PNeu} (a) and (b)  we continue (\ref{Neu1}) and obtain
\begin{equation}
\left( e^{-iH^\epsi{t/\epsi}} - P_1\, \mathcal{U}^*\,e^{-iH^\epsi_{\rm BO}{t/\epsi}}
\,\mathcal{U}
\right)\, P_\Gamma^\alpha \,=
 \,  \left( e^{-iH^\epsi{t/\epsi}} - P_1\, \mathcal{U}^*\, \mathcal{E}\,e^{-iH^\epsi_{\rm BO}{t/\epsi}}
\,\mathcal{U}
\right)\, P_\Gamma^\alpha + O(\epsi)\,.
\end{equation}
We proceed as in the globally isolated band case and write
\begin{eqnarray} \lefteqn{
 \left( e^{-iH^\epsi{t/\epsi}} - P_1\, 
\mathcal{U}^*\, \mathcal{E}\,e^{-iH^\epsi_{\rm BO}{t/\epsi}} \,\mathcal{U}
\right)\, P_\Gamma^\alpha }
\nonumber\\
&=& -i e^{-iH^\epsi{t/\epsi}}
\int_0^{{t/\epsi}}\,ds\,e^{iH^\epsi s}\,\big(H^\epsi\,P_1\,\mathcal{U}^*\,\mathcal{E}\,-
\,P_1\,\mathcal{U}^*\,\mathcal{E}\,
H^\epsi_{\rm BO}\big)e^{-iH^\epsi_{\rm BO}s}\, \mathcal{U}\,P_\Gamma^\alpha \nonumber\\
&=& -i e^{-iH^\epsi{t/\epsi}}
\int_0^{{t/\epsi}}\,ds\,e^{iH^\epsi s}\,\big(H^\epsi\,-\,H^\epsi_{\rm diag})\,P_1 
\,\mathcal{U}^*\, \mathcal{E} e^{-iH^\epsi_{\rm BO}s}\, 
\mathcal{U}\,P_\Gamma^\alpha\,\label{II0}
\\
&&-\, 
 i e^{-iH^\epsi{t/\epsi}}
\int_0^{{t/\epsi}}\,ds\,e^{iH^\epsi s}\,\big(H^\epsi_{\rm diag}\,P_1 
\,\mathcal{U}^*\, \mathcal{E} - 
\,P_1\,\mathcal{U}^*\,\mathcal{E}\,
H^\epsi_{\rm BO}\big)
e^{-iH^\epsi_{\rm BO}s}\, 
\mathcal{U}\,P_\Gamma^\alpha\,,\label{II2}
\end{eqnarray}
where 
\[
H^\epsi_{\rm diag} := P_3\, H^\epsi\,P_3\,.
\]
One can now show that  (\ref{II0})
is bounded in norm by a constant times $\epsi(1+|t|)$ 
using exactly the same sequence of arguments as in the proof in
the previous section. One must only keep track of the ``hierarchy'' of smoothed projections, e.g., 
instead of (\ref{HD1}) one has 
\[
\big( H^\epsi\,-\,H^\epsi_{\rm diag} \big)P_1 = (1-P_3) 
\left[- \frac{\epsi^2}{2}\Delta_X\otimes{\bf 1}, P_2\right] P_1 \,+\,O(\epsi^2).
\]
The adjoint part drops out completely, because this time only the difference on the band,
i.e.\ on Ran$P_1$, is of interest.
Note also that the smoothed projections $P_i$ are bounded operators on the respective scaled
Sobolev spaces and thus, 
according to Proposition \ref{PNeu} (d), all estimates hold in the norm of $\mathcal{L}(\mathcal{H})$.

It remains to show that also (\ref{II2}) is $O(\epsi)$. First note that, according 
to Proposition \ref{PNeu} (c), commuting $\mathcal{E}$ and $H_{\rm BO}^\epsi$ yields an error of 
order $O(\epsi^2)$ in the integrand and thus an error of order $O(\epsi)$ after integration.
For $\phi\in W^2$ we compute
\begin{eqnarray}
(H_{\rm diag}^\epsi\,P_1\,\mathcal{U}^*\phi)(X) &= &
\1I_1(X) \,E(X)\,\phi(X)\chi(X)  +
\1I_1(X) \left( \frac{\epsi^2}{2} \big(-i\nabla_X + A_{\rm ext}\big)^2  \,\phi\right) (X)
\,\chi(X)\nonumber\\
&& +\,\epsi\,
\1I_1(X) \left( -i\epsi \nabla \phi\right) (X) \cdot \left(-
 i \langle \chi(X), \nabla_X \chi (X)\rangle_{\mathcal{H}_{\rm e}}\right) \,\chi(X)\nonumber\\
&& - \,i\,\epsi\, (\nabla\1I_1)(X)\cdot\left( -i\epsi \nabla \phi\right)(X)
 \,\chi(X)\, +\,O(\epsi^2)\,.
\end{eqnarray}
On the other hand, again for $\phi\in W^2$,
\begin{eqnarray}
(P_1\, \mathcal{U}^*\,
H^\epsi_{\rm BO}\,\phi)(X) &= &
\1I_1(X) \,E(X)\,\phi(X)\chi(X)  +
\1I_1(X)
 \left( \frac{\epsi^2}{2} \big(-i\nabla_X + A_{\rm ext}\big)^2  \,\phi\right) (X) \,\chi(X)\nonumber\\
&& +\,\epsi\,
\1I_1(X) \left( -i\epsi \nabla \phi\right) (X) \cdot 
 A_{\rm geo}(X)\,\chi(X)\, +\,O(\epsi^2)\,.
\end{eqnarray}
Hence 
\[
H_{\rm diag}^\epsi\,P_1\,\mathcal{U}^*\mathcal{E} - P_1\, \mathcal{U}^*\,
H^\epsi_{\rm BO}\,\mathcal{E} = -\epsi\,\mathcal{U}^*\,
 (\nabla\1I_1)\cdot\epsi\nabla_X\,\mathcal{E} + O(\epsi^2)
\]
Thus the norm of (\ref{II2}) is, up to an error of order $O(\epsi)$, bounded by the norm of
\begin{equation} \label{FI}
\epsi\,\mathcal{U}^*\,\int_0^{t/\epsi}\,ds\, (\nabla\1I_1)\cdot\epsi\nabla_X\,\mathcal{E}\,
e^{-iH_{\rm BO}^\epsi s}\,\mathcal{U}\,P_\Gamma^\alpha\,.
\end{equation}
$(\nabla\1I_1)\cdot\epsi\nabla_X\,\mathcal{E}$ is a bounded operator and we can apply
Proposition \ref{CLPROP} in the integrand of (\ref{FI}) once more, this time however with
the smoothed projection $P_0$, and obtain
\begin{equation} \label{FFI}
(\ref{FI}) \,=\,\epsi\,\mathcal{U}^*\,\int_0^{t/\epsi}\,ds\, 
(\nabla\1I_1)\cdot\epsi\nabla_X\,\mathcal{E}\,\1I_0\,
e^{-iH_{\rm BO}^\epsi s}\,\mathcal{U}\,P_\Gamma^\alpha\,+O(\epsi)=O(\epsi).
\end{equation}
The last equality in (\ref{FFI}) follows from the fact that 
$[\epsi\nabla_X\,\mathcal{E},\1I_0]=O(\epsi)$
 and that $(\nabla\1I_1)$ and $\1I_0$ are disjointly supported.
\medskip

{\em Proof of Proposition \ref{MT2}.}\quad For the following calculations we continue 
$\chi(\cdot)\in C_{\rm b}^\infty(\Lambda,\mathcal{H}_{\rm e})$ arbitrarily to a function
$\chi(\cdot)\in C_{\rm b}^\infty(\IR^n,\mathcal{H}_{\rm e})$ by possibly modifying it on
$\Lambda\setminus(\Lambda-\delta/2)$.
For $\phi$ in a dense subset of $L^2(\Lambda-\delta)$
and $X\in\Lambda-\delta/2$, by making the substitutions  $\widetilde k = \epsi k$ and 
$\widetilde Y = (Y-X)/\epsi$ and using Taylor expansion with rest,
we have:
\begin{eqnarray}
\lefteqn{\hspace{-1cm}
\left(\big( a^{\rm W,\epsi} \otimes {\bf 1}\big)\,\phi\chi\right)(X) =
(2\pi)^{-n}\,\int\,dY\,dk\,a\left(\frac{X+Y}{2},\epsi k\right)
e^{-i(X-Y)\cdot k}\,\phi(Y)\chi(Y)}\nonumber\\
&=&
(2\pi)^{-n}\,\int\,d\widetilde Y\,\widehat a^{(2)}\left(X+\frac{\epsi}{2}\widetilde Y,
-\widetilde Y\right)\,\phi(X+\epsi\widetilde Y)\,\chi(X)
\nonumber\\
&&+\,\epsi\,
(2\pi)^{-n}\,\int\,d\widetilde Y\,\widehat a^{(2)}\left(X+\frac{\epsi}{2}\widetilde Y,
-\widetilde Y\right)\,\phi(X+\epsi\widetilde Y)\,\widetilde Y\cdot
\big(\nabla_X \chi\big)
(f(X,\epsi\widetilde Y))\nonumber\\
&=&
\left(\mathcal{U}^*\,a^{\rm W,\epsi}\,\mathcal{U}\,\phi\chi\right)(X)\,
+ \, R^\epsi\,.\label{blabla}
\end{eqnarray}
From (\ref{blabla}) we conclude that
\begin{equation}
\left\| \big(
\1I_{\Lambda-\delta/2}(\cdot)\otimes{\bf 1} \big)\,\left( a^{\rm W,\epsi} \otimes {\bf 1} \, - \, 
\mathcal{U}^*\,a^{\rm W,\epsi} \,\mathcal{U} \right) P_{\Lambda -\delta} \right\|
\leq \,\|R^\epsi\|\,.
\end{equation}
Since 
\begin{eqnarray*}\lefteqn{
\left\| \big({\bf 1}-
\1I_{\Lambda-\delta/2}(\cdot)\otimes{\bf 1} \big)\,\left( a^{\rm W,\epsi} \otimes {\bf 1} \, - \, 
\mathcal{U}^*\,a^{\rm W,\epsi} \,\mathcal{U} \right) P_{\Lambda -\delta} \right\|
}\\&=&
\left\| \big({\bf 1}-
\1I_{\Lambda-\delta/2}(\cdot)\otimes{\bf 1} \big)\,\left( a^{\rm W,\epsi} \otimes {\bf 1} \, - \, 
\mathcal{U}^*\,a^{\rm W,\epsi} \,\mathcal{U} \right)  \big(
\1I_{\Lambda-\delta}(\cdot)\otimes{\bf 1} \big) P_{\Lambda -\delta} \right\|= O(\epsi^n)
\end{eqnarray*}
for arbitrary $n$, Proposition \ref{MT2} follows by showing 
 that $R^\epsi$ is of order $\epsi$:
\begin{eqnarray*}
\|R^\epsi\| &\leq&\,\epsi\,
(2\pi)^{-n}\,\int \,d\widetilde Y\,\left\|
\widehat a^{(2)}\left(\cdot+\frac{\epsi}{2}\widetilde Y,
-\widetilde Y\right)\,\phi(\cdot+\epsi\widetilde Y)\,\widetilde Y
\cdot\big(\nabla_X \chi\big)
(f(\cdot,\epsi\widetilde Y))\right\|_\mathcal{H}
\\
&\leq&
\epsi\,(2\pi)^{-n}\,\sup_{X\in\IR^n}\|(\nabla_X\chi)(X)\|_{\mathcal{H}_{\rm e}}
\int \,d\widetilde Y\,\left\|
\widehat a^{(2)}\left(\cdot+\frac{\epsi}{2}\widetilde Y,
-\widetilde Y\right)\,|\widetilde Y|\,\phi(\cdot+\epsi\widetilde Y)\right\|_{L^2(\IR^n)}
\\
&\leq&
\epsi\,C\,\|\phi\|_{L^2(\IR^n)}
 \int\,d\widetilde Y\, \sup_{X\in \IR^n}\,|\widetilde Y|\,| \widehat  a^{(2)} (X,
\widetilde Y)|\\
&=&
\epsi\,\widetilde C\,\|\phi\chi\|_{\mathcal H}\,.
\end{eqnarray*}

\noindent{\bf Acknowledgment:} We are grateful to Andr\'e Martinez and Gheorghe Nenciu for explaining to
us their work in great detail. S.\ T.\  would like to thank 
George Hagedorn for stimulating discussions and, in particular, for helpful advice on
questions concerning the Berry connection  and Markus Klein and Ruedi Seiler for
explaining their treatment of Coulomb singularities. 
We thank Caroline Lasser and Gianluca Panati for careful reading of the manuscript and the referee for pointing out Reference 
 \cite{HerrinHowland}.

\end{document}